\documentclass[aps,prb,twocolumn,showpacs,superscriptaddress,floatfix]{revtex4-1}

\usepackage[colorlinks=true, citecolor=black, linkcolor=black, urlcolor=black]{hyperref}

\usepackage[all]{hypcap}
\usepackage{amsmath}
\usepackage{enumerate,bm}
\usepackage{graphicx}
\usepackage{grffile} 
\usepackage{color}
\usepackage{esint}
\usepackage{textpos}
\usepackage[usenames,dvipsnames]{xcolor}
\usepackage{subfigure}
\usepackage{flushend}
\usepackage{tabularx}

\usepackage{amssymb}
\usepackage{float}
\usepackage{subfigure}
\usepackage{ulem} 

\newcommand{\kk}{{\bm{k}}}

\newcommand{\tmin}{t_\mathrm{min}}

\begin{document}

\author{M.~A.~Sentef}
\email[]{michael.sentef@mpsd.mpg.de}
\affiliation{HISKP, University of Bonn, Nussallee 14-16, D-53115 Bonn, Germany}
\affiliation{Max Planck Institute for the Structure and Dynamics of Matter,
Center for Free Electron Laser Science, 22761 Hamburg, Germany}
\author{A.~F.~Kemper}
\affiliation{Lawrence Berkeley National Laboratory, 1 Cyclotron Road, Berkeley, CA 94720, USA}
\affiliation{Department of Physics, North Carolina State University, Raleigh, NC 27695, USA}
\author{A.~Georges}
\affiliation{Centre de Physique Th\'eorique, 
\'Ecole Polytechnique, CNRS, 91128 Palaiseau Cedex, France}
\affiliation{Coll\`ege de France, 11 place Marcelin Berthelot, 75005 Paris, France}
\affiliation{Department of Quantum Matter Physics, University of Geneva, 24 Quai Ernest-Ansermet, 1211 Geneva 4, Switzerland}
\author{C.~Kollath}
\affiliation{HISKP, University of Bonn, Nussallee 14-16, D-53115 Bonn, Germany}


\title{
Theory of light-enhanced phonon-mediated superconductivity
}
\date{\today}
\begin{abstract}
We investigate the dynamics of a phonon-mediated superconductor driven out of equilibrium. 
The electronic hopping amplitude is ramped down in time, resulting in an increased electronic density of states. 
The dynamics of the coupled electron-phonon model is investigated by solving Migdal-Eliashberg equations for the 
double-time Keldysh Green's functions. The increase of the density of states near the Fermi level leads to an enhancement of superconductivity when the system thermalizes to the new state at the same temperature. We provide a time- and momentum-resolved view on this thermalization process, and show that it involves fast processes associated with single-particle scattering and much slower dynamics associated with the superconducting order parameter. The importance of electron-phonon coupling for the rapid enhancement and the efficient thermalization of superconductivity is demonstrated, and the results are compared to a BCS time-dependent mean-field approximation.  
\end{abstract}
\maketitle

\section{Introduction}

Light control of structural and electronic properties of solids is a tantalizing prospect of ultrafast materials science \cite{orenstein_ultrafast_2012, zhang_dynamics_2014, forst_mode-selective_2015}. In pump-probe experiments, a short pump laser pulse drives a solid out of equilibrium. The ensuing dynamics is monitored with a second probe pulse at well-defined delay times. Pump excitations at optical frequencies usually create electron-hole excitations, which can be used to study transient dynamics in a variety of correlated materials \cite{orenstein_ultrafast_2012, zhang_dynamics_2014}, like Mott or charge-density wave insulators \cite{perfetti_time_2006, schmitt_transient_2008, hellmann_ultrafast_2010, rohwer_collapse_2011, hellmann_time-domain_2012}, or superconductors \cite{demsar_pair-breaking_2003, gedik_single-quasiparticle_2004, perfetti_ultrafast_2007, graf_nodal_2011, cortes_momentum-resolved_2011, beck_energy-gap_2011, smallwood_tracking_2012}. In contrast, lower frequency mid-IR or THz lasers can excite the system in resonance with structural \cite{rini_control_2007} or other collective modes. In particular, intense THz light pulses enable a mode-selective vibrational excitation \cite{rini_control_2007}, opening up the field of ``nonlinear phononics'' \cite{forst_nonlinear_2011, subedi_theory_2014, forst_mode-selective_2015}. 

A lattice deformation can be induced that lasts for hundreds of femtoseconds \cite{rini_control_2007, forst_nonlinear_2011, forst_driving_2011, subedi_theory_2014}, which has been suggested as a basis for light-enhanced superconducting-like nonequilibrium states \cite{fausti_light-induced_2011, mankowsky_nonlinear_2014, kaiser_optically_2014, hoppner_redistribution_2015, mitrano_possible_2016}. Thus the important question arises how fast the electrons in a solid can follow a nonadiabatic change of the lattice structure. In particular, the situation is unclear for slow collective modes in a symmetry-broken ordered state, such as a superconductor or a charge-density wave.

Theoretically, the order parameter dynamics in purely electronic models has been investigated in BCS mean-field theories for superconductors \cite{volkov_collisionless_1974, barankov_collective_2004, warner_quench_2005, yuzbashyan_nonequilibrium_2005, yuzbashyan_dynamical_2006, barankov_synchronization_2006, yuzbashyan_dynamics_2009, papenkort_coherent_2007, papenkort_coherent_2008, unterhinninghofen_theory_2008, schnyder_resonant_2011, akbari_theory_2013, mansart_coupling_2013, krull_signatures_2014, matsunaga_light-induced_2014, tsuji_theory_2015, peronaci_transient_2015}, and in nonequilibrium dynamical mean-field theory for antiferromagnets \cite{tsuji_nonthermal_2013, tsuji_nonequilibrium_2013}. In contrast to such closed systems, where the electronic energy is conserved after the external perturbation, in electron-lattice systems energy is transferred between electrons and phonons via electron-phonon (el-ph) coupling \cite{allen_theory_1987}. The electronic relaxation in electron-phonon models has been theoretically investigated using a variety of methods \cite{yonemitsu_coupling-dependent_2009, vidmar_nonequilibrium_2011, lee_ultrafast_2011, golez_relaxation_2012, matsueda_relaxation_2011, kemper_mapping_2013, werner_phonon-enhanced_2013, sentef_examining_2013, hohenadler_charge_2013, kemper_effect_2014, baranov_theory_2014, murakami_interaction_2015, dorfner_real-time_2015, sayyad_coexistence_2015, rameau_time-resolved_2015}. 

\begin{figure}[ht!pb]
\includegraphics[clip=true, trim=0 0 0 0,width=\columnwidth]{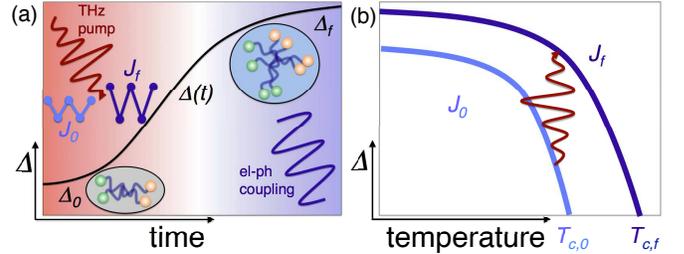}
\caption{
{\bf THz pump enhances superconductivity.} 
(a) Through a lattice distortion, the electronic hopping amplitude decreases from $J_0$ to $J_f$ (red shaded area). As a consequence, the superconducting order parameter $\Delta_0$ is boosted to a larger value $\Delta(t) > \Delta_0$. At longer time scales (blue shaded area) the order parameter approaches its thermal value $\Delta_f$ corresponding to $J_f$ in the presence of efficient electron-phonon (el-ph) coupling. (b) Sketch of equilibrium order parameters $\Delta_0$ and $\Delta_f$ corresponding to $J_0$ and $J_f < J_0$, respectively, leading to a larger critical temperature $T_{c,f} > T_{c,0}$.}
\label{fig1}
\end{figure}

In this work we investigate the nonequilibrium dynamics of a phonon-mediated superconductor induced by a transiently modified electronic structure through nonlinear phonon coupling.  We consider a tight-binding el-ph Hamiltonian which contains both a retarded pairing interaction mediated by phonons as well as dissipation of heat into the lattice. The light-induced lattice distortion is accounted for by a change of the electronic hopping amplitude $J_0$ to a smaller value $J_f$ on a typical time scale of fractions of a picosecond. Due to this change the electronic density of states close to the Fermi surface is enhanced, which results in an increased equilibrium order parameter $\Delta_f$ (see Fig.~\ref{fig1}) in the weak-coupling regime assumed throughout this work.

Out of equilibrium, the order parameter is therefore expected to increase if the change is slow enough and not too much energy is deposited into the electronic degrees of freedom. Since typically the time scale of the lattice distortion -- while much longer than the bare electronic time scale -- is rapid compared to the slow collective dynamics of the superconducting condensate, understanding the response of the superconducting order parameter $\Delta(t)$ to such a relatively fast change is of great importance. 
We show that even for this rapid change of the lattice structure, the superconducting order parameter can be drastically enhanced. The dynamics can be separated into two different regimes: (i) the short time dynamics of the order parameter, which can approximately be described by BCS theory, and (ii) the intermediate to long time dynamics, where el-ph scattering and the relaxation of energy into the dissipative phonon bath dominate. Importantly, the phonon dissipative channel is essential for asymptotically reaching the final thermal value. Surprisingly, very fast nonadiabatic ramps are predicted to lead to quick enhancement of superconductivity on very short time scales in the presence of dissipation.

The paper is organized as follows: Section \ref{sec:model} contains model and methods. In Section \ref{sec:results} the main results are presented. These results are put into context with conclusions and an outlook in Section \ref{sec:conclusions}. The Appendix contains more detailed information about the BCS formalism at finite temperature and additional results on the intermetiate time regime. 
 
\section{Model and Methods} 
\label{sec:model}
 
\subsection{Electron-phonon Hamiltonian}

We investigate the electron-phonon Hamiltonian 
\begin{align}
\mathcal H &= \sum_{\bm{k}\sigma} \epsilon(\kk,t) c^\dagger_{\bm{k}\sigma} c^{}_{\bm{k}\sigma}   + \sum_{\bm{q},\gamma} \Omega_{\gamma} b_{\bm{q},\gamma}^\dagger b^{}_{\bm{q},\gamma} \nonumber\\ &-
		\sum_{\bm{q},\gamma,\sigma} g_{\gamma} c_{\bm{k+q}\sigma}^\dagger c_{\bm{k}\sigma} \left( b_{\bm{q},\gamma} + b_{-\bm{q},\gamma}^\dagger \right)
\end{align}
with fermionic creation operators $c^\dagger_{\bm{k}\sigma}$ for dimensionless momentum $\bm{k}=(k_x,k_y)$ and spin $\sigma$ $=$ $\uparrow,\downarrow$ on a two-dimensional square lattice with dispersion $\epsilon(\kk,t) = -2 J(t) (\cos k_x+\cos k_y)$ at half-filling. This choice of band filling is made for numerical convenience. Away from particle-hole symmetric filling, the chemical potential would have to be adjusted to keep the filling fixed at different temperatures, which we avoid. 

The time dependence of the electronic hopping amplitude $J(t)$ mimics a deformation of the lattice induced via a nonlinear coupling to an IR active optical phonon driven by the THz light pulse~\cite{subedi_theory_2014,knap_dynamical_2015}. Thus, the excited phonon is treated classically. We assume for $t<\tau$ a linear ramp  $J(t) = J_0 + (J_f - J_0) \frac{t}{\tau}$ and for $t>\tau$ the constant $J(t) = J_f$ with $J_0 =$ 0.25 eV, $J_f =$ 0.20 eV, and ramp time $\tau$. The change of the hopping parameter by 20\% is rather large, but not out of reach for an experimental realization \cite{subedi_theory_2014,mitrano_possible_2016,limelette_mott_2003}. 
Furthermore, in an experiment the deformation of the lattice typically lasts for several picoseconds, and we focus on the dynamics within this time frame. Energies are measured in eV, and time scales in fs, using $\hbar = 0.658$ eV$\times$fs. 

The electrons are coupled to branches ($\gamma$) of phonons with bosonic creation operators $b_{\bm{q},\gamma}^\dagger$, energy $\Omega_{\gamma}$, and electron-phonon coupling $g_{\gamma}$. These quantum phonons model the different relaxation channels present in the material and should not be confused with the externally excited phonon mentioned above. We consider a dominant optical phonon at $\Omega_{\text{opt}} =$ 0.1 eV, which induces superconductivity, and a continuum of acoustic low-energy phonons. The distribution of the acoustic phonons is given later. We use a reference set of electron-phonon couplings, labeled by set (i), and another set with the same spectrum but reduced coupling strengths labeled by set (ii). The parameters used for the different sets are listed in Table~\ref{phonon_table}.  We solve this model in the Migdal-Eliashberg approximation \cite{eliashberg_interactions_1960, migdal_interaction_1958, kemper_direct_2015} with a local, self-consistent self-energy for the electrons, and treat the phonons as an infinite heat bath at equilibrium. The effective phonon spectra weighted by el-ph coupling (Eliashberg functions) for case (i) and a parameter set without acoustic branch are shown in the inset to Fig.~\ref{fig4}(a). 

The important difference between BCS mean-field treatments, possibly including phenomenological damping, and the present approach is that we explicitly treat a double-time self-energy with nonzero imaginary part, which involves true correlation and memory effects and accounts for a frequency structure of the phonon spectrum. This will be shown to be important in this work in the context of absence or presence of low-energy acoustic phonons. From the point of view of superconducting pairing, the explicit treatment of the retarded self-energy also gives a frequency structure to the anomalous self-energy and thus defines a natural energy cutoff to pairing while allowing for a fully gauge-invariant theory. Such a cutoff can be introduced in BCS-like theories as well, but is often neglected because it violates gauge invariance.

\subsection{Time evolution}

The time evolution is obtained from solutions of the Kadanoff-Baym-Gor'kov equations \cite{stan_time_2009, kemper_direct_2015} in the Keldysh Green function formalism, described in detail below. We choose initial conditions that put the system in the superconducting initial state below $T_c$. We ignore the competing instability towards charge-density wave order at half-filling, which is always possible within a mean-field scheme. The time-dependent order parameter $\Delta(t)$ is defined by 
\begin{align}
\frac{\Delta(t)}{\Delta_0} &\equiv \frac{\sum_\kk f_\kk(t)}{\sum_\kk f_{\kk}(0)} 
\end{align}
using the dimensionless momentum-resolved anomalous expectation value $f_\kk(t) \equiv F^<_\kk(t,t) \equiv \langle c_{-\kk\downarrow}(t) c_{\kk\uparrow}(t) \rangle$. The initial value $\Delta_0=\Delta(t=0)$ and final value $\Delta_f$ are obtained from the anomalous component of the equilibrium self-energy, including energy band renormalization with quasiparticle weight $Z$ due to el-ph coupling (see below).

Our choice of a large el-ph coupling $\lambda$, which results in a large value of the order parameter compared to real materials \footnote{For conventional superconductors, $\Delta_0$ is at most in the few meV range, for example up to 7 meV for the larger gap in MgB$_2$ \cite{choi_origin_2002}. This is roughly a factor of three smaller than the zero-temperature limit of our model system.}, is motivated by the times we can reach in the numerical simulations. Even though the Migdal-Eliashberg approximation is not expected to be quantitatively accurate in this regime, the generic effects observed should remain valid. 

\subsection{Kadanoff-Baym-Gor'kov equations}

We employ the Kadanoff-Baym-Gor'kov formalism and its application to the superconducting state in the el-ph model as described in Ref.~\onlinecite{kemper_direct_2015}. We utilize the standard two-time Keldysh formalism\cite{stan_time_2009}, where the 
contour Green functions are 2x2 matrices in Nambu space,
\begin{align}
\bar{G}^\mathcal{C}_\kk(t,t') &= -i\left\langle\mathcal{T_C}
\left(
\begin{array}{cc}
c_{\kk\uparrow}(t) c^\dagger_{\kk\uparrow}(t') & c_{\kk\uparrow}(t) c_{-\kk\downarrow}(t') \\
c^\dagger_{-\kk\downarrow}(t) c^\dagger_{\kk\uparrow}(t')  & c^\dagger_{-\kk\downarrow}(t)  c_{-\kk\downarrow}(t')
\end{array}
\right)\right\rangle \\
&\equiv
\left(
\begin{array}{cc}
G^\mathcal{C}_{\kk,\uparrow}(t,t') & F^\mathcal{C}_\kk(t,t') \\ 
F^{\dagger\mathcal{C}}_\kk(t,t')  & -G^\mathcal{C}_{-\kk,\downarrow}(t',t) 
\end{array}
\right),
\end{align}
where $t$ and $t'$ lie on the Keldysh contour, and $\mathcal{T_C}$ is the contour time-ordering operator. In the following, we use units with $\hbar \equiv k_B \equiv 1$. 

The matrix equations of motion with a contour self-energy $\bar\Sigma^\mathcal{C}$ to be specified below are 
\begin{align}
\left(i\partial_t \bar\tau_0 - \bar\epsilon_\kk(t) \right) \bar{G}^\mathcal{C}_\kk(t,t') 
&= \delta^\mathcal{C}(t,t') \bar\tau_0 + \nonumber\\ &\int_\mathcal{C}dz\  \bar\Sigma^\mathcal{C}(t,z) \bar G_\kk^\mathcal{C}(z,t'),
\end{align}
with
\begin{align}
\bar\epsilon_\kk(t) &= \left(
\begin{array}{cc}
\epsilon_\uparrow(\kk, t) & 0 \\
0 & -\epsilon_\downarrow(-\kk, t)
\end{array}
\right)
\end{align}
where $\bar\tau_0$ is the identity matrix.

On the Keldysh contour, the Langreth rules can be applied to separate the contour equation into the following components: the Matsubara ($M$), lesser ($<$), and greater ($>$) Green functions, as well as the mixed real-imaginary $\rceil/\lceil$ Green functions. The various components can be transformed or combined into others via the relations
\begin{align}
\bar G^\lessgtr(t,t')^\dagger &= -\bar G^\lessgtr(t',t) \\
\bar G^\lceil(-i\tau,t)^\dagger &= \bar G^\rceil(-i(\beta-\tau),t).
\end{align}
The equations of motion, letting the contour start at $\tmin$, are 
\begin{widetext}
\begin{subequations}
\begin{align}
\big[ -\partial_\tau \bar\tau_0 - \bar\epsilon_\kk(\tmin)\big] \bar G_\kk^M(\tau) =& i\delta(\tau)\bar\tau_0 - i \int_0^\beta dz\ \bar\Sigma^M(\tau-z) \bar G_\kk^M(z), \\
\big[ i\partial_t \bar\tau_0- \bar\epsilon_\kk(t) \big] G_\kk^\rceil(t,-i\tau) =& \int_{\tmin}^t dz\ \bar \Sigma^R(t,z) \bar G_\kk^\rceil(z, -i\tau) 
-i \int_0^\beta dz\ \bar\Sigma^\rceil(t,-iz) \bar G_\kk^M(z -\tau),\\
\big[ i\partial_t \bar\tau_0- \bar\epsilon_\kk(t) \big] \bar G_\kk^\gtrless(t,t') =& \int_{\tmin}^t dz\  \bar\Sigma^R(t,z) \bar G_\kk^\gtrless(z, t') 
+\int_{\tmin}^{t'} dz\  \bar\Sigma^\gtrless(t, z) \bar G_\kk^A(z, t') 
 - i \int_0^\beta dz\ \bar\Sigma^\rceil(t,-iz) \bar G_\kk^\lceil(-iz,t'),
 \label{eq:Gless}
\end{align}
\label{eq:eoms}
\end{subequations}
\end{widetext}
These equations are solved on the contour by using massively parallel computation and a time stepping algorithm for integro-differential equations as described in Ref.~\onlinecite{stan_time_2009} and numerical details are given in Sec.~\ref{sec:num}.

Note that we can also drop the spin index $\uparrow, \downarrow$ on the normal components of the Nambu Green function, since in the absence of magnetic order we have $G^\mathcal{C}_{\kk,\uparrow}(t,t') = G^\mathcal{C}_{\kk,\downarrow}(t,t') \equiv G^\mathcal{C}_{\kk}(t,t')$. This relation is used for the calculation of the momentum-resolved normal and anomalous densities
\begin{align}
n_\kk(t) &\equiv -i G_{\kk}^<(t,t),\\
f_\kk(t) &\equiv -i F_{\kk}^<(t,t).
\end{align}

\subsection{Migdal-Eliashberg approximation to electron-phonon coupling}

In this work, we employ the Migdal-Eliashberg approximation to electron-phonon coupling of electrons to the phononic relaxation channels which are explicitly treated in our calculation. These phonons should not be confused with the classical phonons involved in the THz driving and nonlinear phonon excitation processes.
 We use a perturbative treatment of the electronic self-energy at the self-consistent Born level
\begin{align}
\bar\Sigma^\mathcal{C}(t,t') = i \int d\Omega \; \alpha^2 F(\Omega) \; \bar\tau_3\ \bar G_\mathrm{loc}^\mathcal{C}(t,t') \bar\tau_3\ D^\mathcal{C}_0(\Omega,t,t').
\end{align}
Here $\bar\tau_3$ is the $z$ Pauli matrix in Nambu space,
and $\bar G^\mathcal{C}_\mathrm{loc}(t,t') = \sum_{\bm{k}} \bar G^\mathcal{C}_{\bm{k}}(t,t')$ the local Green function.

The quantum phonons  are kept at fixed equilibrium temperature neglecting the phonon self-energy. The Keldysh propagator for a single phonon mode at energy $\Omega$ is given by
\begin{align}
D^\mathcal{C}_0(\Omega,t,t') =& -i \big[ n_B(\beta \Omega) + 1 - \Theta_\mathcal{C}(t,t') \big] e^{i\Omega(t-t')} \nonumber \\
& -i \big[ n_B(\beta \Omega) + \Theta_\mathcal{C}(t,t') \big] e^{-i\Omega(t-t')},
\end{align}
where $n_B(x)$ is the Bose function $n_B(x) = \left[e^x-1\right]^{-1}$, $\beta \equiv (k_B T)^{-1}$ the inverse temperature, and $\Theta_\mathcal{C}(t,t')$ is the contour Heaviside function.

The relevant phonon spectrum in Migdal-Eliashberg theory is the Eliashberg function \cite{allen_theory_1987}, defined here for the case of a spectrum of local phonon modes
\begin{align}
\alpha^2 F(\Omega) &= \sum_{\gamma} \left| g_{\gamma} \right|^2 \delta(\Omega-\Omega_{\gamma}).
\label{eq:a2Fdefinition}
\end{align}

In practice, we use a model function 
\begin{align}
\alpha^2 F(\Omega) =& \alpha^2 F_{\text{opt}}(\Omega) + \alpha^2 F_{\text{acou}}(\Omega),
\end{align}
with an optical branch modelled by a Lorentzian
\begin{align}
\alpha^2 F_{\text{opt}}(\Omega) &= g^2_{\text{opt}} \frac{\delta_{\text{opt}}}{\pi ((\Omega-\Omega_{\text{opt}})^2 + \delta^2_{\text{opt}})}
\end{align}
and an acoustic branch with proper $\Omega^2$ behavior at low energy and a cutoff at $2\Omega_{\text{acou}}$ modelled by 
\begin{align}
\alpha^2 F_{\text{acou}}(\Omega) &= \frac{g^2_{\text{acou}}}{\Omega_{\text{acou}}} \sin^2\left(\frac{\pi\Omega}{2\Omega_{\text{acou}}}\right) \Theta(2\Omega_{\text{acou}}-\Omega).
\label{eq:acou}
\end{align}

Finally, we estimate a dimensionless electron-phonon coupling parameter 
\begin{align}
\lambda &\equiv -\frac{\partial \text{Re} \Sigma^{R}_{11}(\omega)}{\partial\omega} \Big|_{\omega=0} 
\approx -\frac{\text{Im} \Sigma^{M}_{11}(i\omega_0)}{\omega_0},
\end{align}
where $\Sigma^{M}_{11}$ is the normal component of the Matsubara self-energy
\begin{align}
\bar \Sigma(i \omega_n) \equiv -i \int_0^{\beta} d\tau \; e^{i \omega_n \tau} \bar \Sigma(\tau)
\end{align}
with imaginary frequency $i \omega_n = i (2n+1) \pi/\beta$.

The renormalized quasiparticle weight
\begin{align}
Z = \frac{1}{1+\lambda}
\end{align}
reflects the effective mass change near $E_F$ induced by el-ph coupling.
This renormalization is taken into account in computing the Bogoliubov dispersions used in Figs.~3 and 4, and for the initial equilibrium order parameter
\begin{align}
\Delta_0 =  Z \Sigma^{M}_{12}(i \omega_0),
\end{align}
where $\Sigma^{M}_{12}$ is the anomalous component of the Matsubara self-energy.

The individual contributions from the phonon modes are estimated via 
\begin{align}
\lambda_{\text{opt}} \equiv 2 \bar{N}(E_F) \int_0^{\infty} d\Omega \;  \frac{\alpha^2 F_{\text{opt}}(\Omega)}{\Omega},
\end{align}
and equivalently for $\lambda_{\text{acou}}$. Here we use the average electronic density of states $\bar{N}(E_F)$ in a window $\pm\Omega_{\text{opt}}$ around the Fermi level at the initial equilibrium. This expression is only strictly valid in the weak coupling limit, where the self-energy contributions from optical and acoustic branches add up to the total self-energy.

The detailed parameters used for the runs in this paper are displayed in Table \ref{phonon_table}. We note that the values of $\lambda$ used in this work are relatively high in order for weak-coupling perturbation theory to be a quantitatively accurate approximation. This choice is motivated by numerical feasibility. In order to see a crossover from nonadiabatic towards adiabatic behavior, the time scale associated with the initial order parameter should not be too large compared to the time scales we are able to reach in the simulation. A rather large value of $\lambda$ guarantees that we have a sizeable order parameter as well as relatively short relaxation times. We expect that using a more moderate value of $\lambda$ would change the quantitative results, but not the conclusions drawn from our work.

A few words are in order regarding the choice of the heat bath approximation for the phonons, leaving them at their initial thermal equilibrium. This can be justified by three arguments: (i) The THz-induced lattice modification leading to a change of the electronic hopping amplitude is a subtle excitation of the electrons, compared to the immediate optical excitation of electron-hole pairs with higher energy photons. Thus, the excess energy in the electronic subsystem is relatively small. (ii) We use a continuous spectrum of phonons rather than just a single sharp mode. The heat capacity of this spectrum of phonons is supposedly large, hence the heating of the phonons via the transfer of the small amount of excess energy is practically negligible. (iii) In reality, if there is heating of the phonons, the electrons will thermalize towards the enhanced lattice temperature -- at least within an effective temperature description -- before the excess heat is transferred from the irradiated sample surface to the bulk crystal. The heating of the lattice strongly depends on the sample and the precise experimental conditions, effects that are beyond the scope of our present model study. 

\begin{table}
\begin{ruledtabular}
\begin{tabular}{c | c | c | c}
Parameter set & (i) & (ii) & w/o acoustic \\ \hline
$\Omega_{\text{opt}}$ [eV] & 0.100 & 0.100 & 0.100 \\ 
$g^2_{\text{opt}}$ [eV$^2$] & 0.040 & 0.032 & 0.052 \\
$\delta_{\text{opt}}$ [eV] & 0.001 & 0.001 & 0.001 \\
$\lambda_{\text{opt}}$ & 0.64 & 0.51 & 0.83 \\
$\Omega_{\text{acou}}$ [eV] & 0.050 & 0.050 & -- \\
$g^2_{\text{acou}}$ [eV$^2$] & 0.025 & 0.020 & -- \\
$\lambda_{\text{acou}}$ & 1.08 & 0.86 & -- \\
\end{tabular}
\end{ruledtabular}
\caption{\label{phonon_table} Parameters for phonon spectra and electron-phonon coupling parameters used in the paper. 
The parameter set without acoustic mode, used for the comparison in Fig.~\ref{fig4}, has larger coupling strength than set (i) in order to match the resulting critical temperature and value of $\Delta_0$.}
\end{table}

\subsection{Numerical details}
\label{sec:num}
The two-dimensional Brillouin zone was discretized with a numerical grid of 80 $\times$ 80 momentum points. Calculations were performed on a reduced $1/8$ zone with a total of 820 momenta. The Kadanoff-Baym-Keldysh contour was discretized, with step sizes of $\delta_\tau =$ 0.1 $\hbar$ (eV)$^{-1}$ and $\delta_t \approx$ 0.21 $\hbar$ (eV)$^{-1}$ $\approx$ 0.14 fs for imaginary and real times axes, respectively. This choice results for example in 1200 imaginary time points and 2800 real time points for the lowest temperature run at $\beta =$ 120 (eV)$^{-1}$ ($T = $ 96 K), implying a total numerical grid of size 6800 $\times$ 6800 for a full double-time Keldysh contour Green function or self-energy. We checked that the numerical results were sufficiently converged as a function of time step size by adding runs with larger step sizes and extrapolating to zero step. Importantly, we used a 3rd order Adams-Bashforth scheme for the numerical integration of the differential equations together with a 6th order Gregory integration for the numerical integrals. The self-consistency cycle for the self-energy typically required a maximum of 5 iterations at each time step, using a standard predictor-corrector scheme \cite{stan_time_2009}. Runs for the electron-phonon simulations typically required 40,000 CPU hours.

\section{Results} 
\label{sec:results}

\begin{figure}[ht!pb]
\includegraphics[clip=true, trim=0 0 0 0,width=\columnwidth]{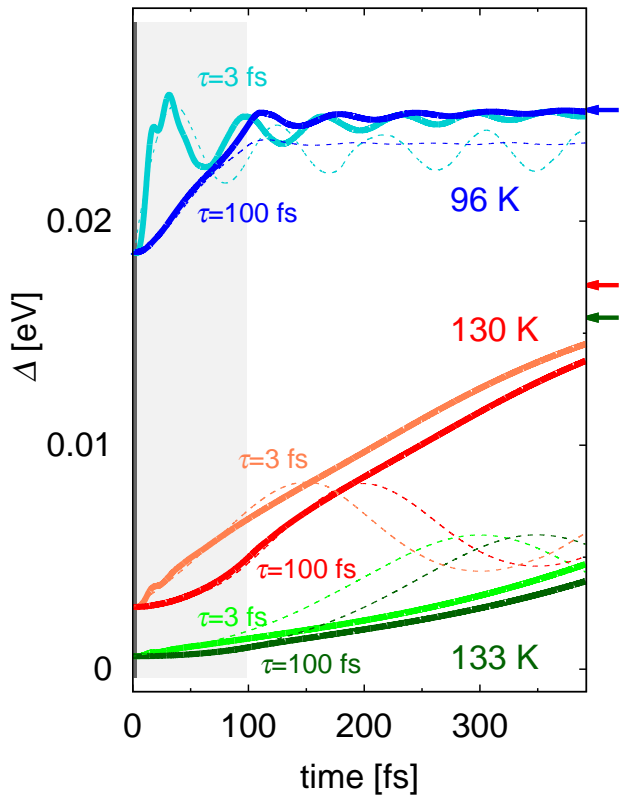}
\caption{
{\bf Light-enhanced superconductivity.} Dynamics during and after $\tau = $ 100 fs (dark colors) and $\tau = $ 3 fs (light colors) ramps for different initial equilibrium temperatures ($T_{c,0} \approx 135$ K). Solid (dashed) lines show the results of the el-ph (BCS) model and arrows indicate the final thermal equilibrium values $\Delta_f$. Dark (light) grey shaded rectangles indicate the ramp durations.}
\label{fig2}
\end{figure}

\begin{figure}[ht!pb]
\includegraphics[clip=true, trim=0 0 0 0,width=0.8\columnwidth]{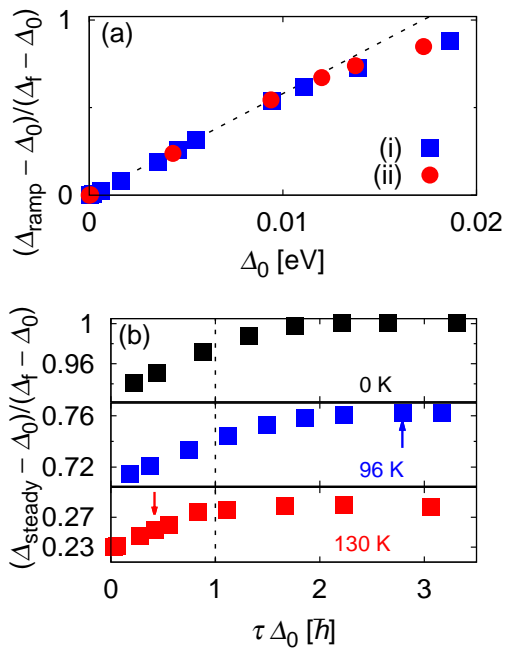}
\caption{
{\bf Initial state and ramp duration dependence.} (a) Order parameter change during the 100 fs ramp, $\Delta_{\text{ramp}}-\Delta_0$, relative to $\Delta_f-\Delta_0$ (symbols). Here $\Delta_{\text{ramp}}$ denotes the order parameter at the end of the ramp, i.e.~at the time $t=\tau=100$fs. The change scales almost linearly with the initial value $\Delta_0$ (dashed line). (b) Dependence of final steady state value on ramp duration $\tau$ within BCS theory for different temperatures. Arrows show the data points for 100 fs ramps corresponding to Fig.~\ref{fig2}. Dashed line indicates $\tau \Delta_0 = \hbar$.
}
\label{fig2bc}
\end{figure}


The fundamental question we address is whether and on which time scales superconducting order can be enhanced by the change of the hopping amplitude. Fig.~\ref{fig2} shows the dynamics of $\Delta(t)$ at different initial temperatures for two different ramp times. For all situations a drastic enhancement of the order parameter is found for sizeable values of $\Delta_0$. The initial increase is much faster for the short ramp duration of 3 fs compared to the slower ramp duration of 100 fs. After the ramp, the order parameter continues to increase for both ramp durations before it slowly approaches the thermal value $\Delta_f$. Damped ``Higgs'' amplitude mode oscillations are observed in some cases, as discussed in detail in Ref.~\onlinecite{kemper_direct_2015}. 

Importantly, the achieved enhancement of superconductivity at short and intermediate times depends crucially on the initial order parameter and the distance from $\Delta_f$. To demonstrate the systematics, Fig.~\ref{fig2bc}(a) shows the fraction of order parameter change during the $\tau=100$ fs ramp, $(\Delta_{\text{ramp}}-\Delta_0) / (\Delta_f-\Delta_0)$, where $\Delta_{\text{ramp}} \equiv \Delta(t=\tau )$. The dashed line highlights the approximate linear dependence of the achieved change on $\Delta_0$ at small $\Delta_0$. In other words, the smaller the initial order parameter, the longer it takes to enhance superconductivity.

In order to gain a deeper understanding of the different regimes of the dynamics, we compare our results to simulations of a BCS model (see Appendix \ref{app:bcs}) with parameters chosen to match $\Delta_0$ and $\Delta_f$ in Fig.~\ref{fig2}. The BCS model contains the electronic dynamics at a mean-field level including the phononic action only as an effective pairing interaction between the electrons. Thus, the BCS model is not expected to be able to reproduce the full dynamics of the el-ph model. The purpose of the comparison between BCS and full el-ph dynamics is to illustrate the importance of dissipation of energy into the phononic bath.

The BCS dynamics captures the main features of the initial increase of the order parameter in agreement with the el-ph model. However, small deviations occur in particular for fast ramps, and the BCS model completely fails to account for the full thermalization on longer time scales. The reasonable agreement at initial times demonstrates that the dynamics of the order parameter at initial times is dominated by the change of the coherence factor relating the bare electrons to the Bogoliubov quasiparticles. In this regime, the main role of the phonons is to generate an effective attractive interaction between the electrons. 

The strong dependence of the order parameter dynamics on the ramp duration is already visible within BCS theory. Whereas the initial increase is strongly accelerated for shorter ramp durations, the reachable steady state value \footnote{A steady state value is extracted using fits with a constant $\Delta_{\text{steady}}$ plus damped oscillations to $\Delta(t)$ for the BCS results after the ramp.} increases with longer ramp durations, 
as shown in Fig.~\ref{fig2bc}(b), due to the reduced heating of the electrons and to the fact that slower ramps are less efficient at breaking Cooper pairs.  The time scale to which the ramp duration has to be compared is $\hbar/\Delta_0$, around which the most important increase of the steady value takes place. The saturation value for long ramp durations can be far from the thermal value at the temperature of the phononic bath. This failure to reach the thermal value is expected due to the presence of conservation laws within the integrable BCS model \cite{barankov_collective_2004, warner_quench_2005, yuzbashyan_nonequilibrium_2005, yuzbashyan_dynamical_2006, barankov_synchronization_2006, tsuji_theory_2015}.  The exception is at $T=0$ (top panel in Fig.~\ref{fig2bc}(b)), where the order parameter follows closely the ground state value in the $\tau \Delta_0 \rightarrow \infty$ limit even within the BCS approximation.

In contrast, the full el-ph model exhibits a very distinct behavior at intermediate and long times. The heat created in the electrons during the ramp is transferred to the phononic bath. As a consequence, the order parameter at long times reaches the expected thermal value independent of the ramp duration. Surprisingly, as seen in Fig.~\ref{fig2}, the phonon dissipative channel even allows to stabilize the strong initial increase directly after a rapid ramp. Thus fast ramps are more favorable than slow ramps with respect to light-induced superconductivity.

\begin{figure}[ht!]
\includegraphics[clip=true, trim=0 0 0 0,width=\columnwidth]{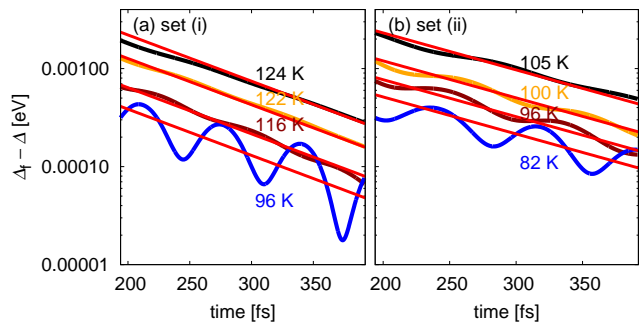}
\caption{
{\bf Thermalization at long times via el-ph coupling.} 
(a) Deviations of order parameter from respective final thermal values for various temperatures and full el-ph coupling parameter set (i) on a logarithmic scale. The approach to the final thermal value is well described by exponential decays (red lines). (b) Same as in (a) but for reduced el-ph coupling parameter set (ii). The slope in the exponential decays is indeed smaller by a factor of 0.8 as expected from the ratios of the electron-phonon couplings. Hence thermalization takes longer for smaller el-ph coupling. 
}
\label{fig_SM2}
\end{figure}

\begin{figure*}[ht!pb]
\includegraphics[clip=true, trim=0 0 0 0,width=\textwidth]{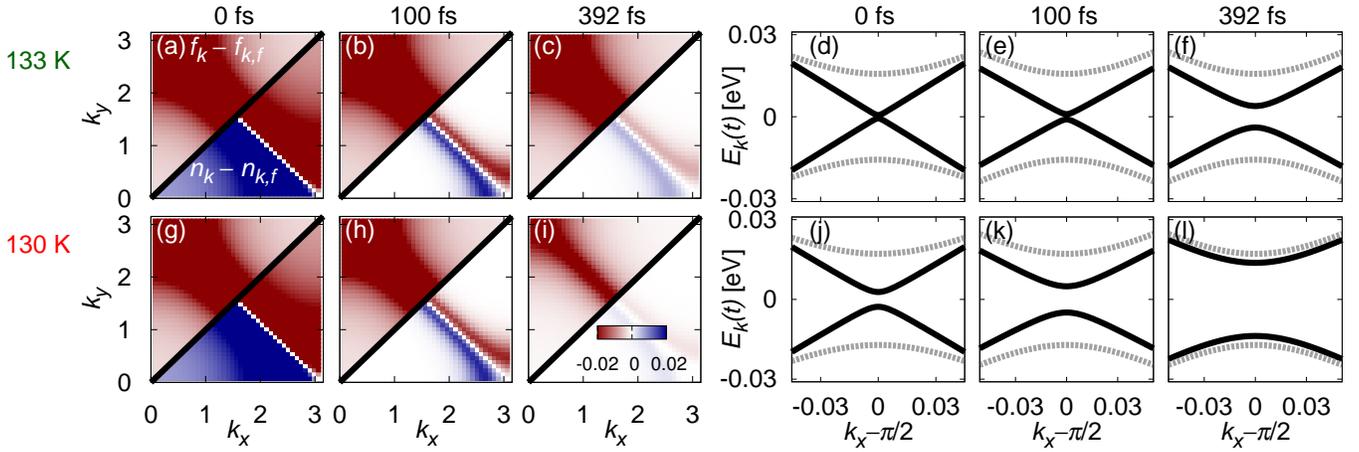}
\caption{
{\bf Time- and momentum-resolved dynamics.} Snapshots of the momentum-resolved deviations from final thermal equilibrium values of the normal and anomalous occupations at times as indicated above, shown in 1/8 Brillouin zone each (separated by black lines) for a case of small $\Delta_0$ (a-c) and intermediate $\Delta_0$ (g-i). Effective Bogoliubov dispersions $E_\kk(t)$ at the corresponding times along a $k_x = k_y$ momentum cut, compared to final thermal dispersions $E_{\kk,f}$ (dashed curves) for the small (d-f) and large (j-l) $\Delta_0$. 
}
\label{fig3}
\end{figure*}

Fig.~\ref{fig_SM2} shows the exponential decay of the order parameter deviation from the final thermal values at long times on a logarithmic scale. Red lines indicate the slopes corresponding to exponentially decaying behavior. As expected the long-time relaxation, which is enabled by el-ph coupling in particular to the acoustic phonon branch, crucially depends on the el-ph coupling strength. Therefore we compare parameter sets at different temperatures for full (a) and reduced (b) el-ph coupling. Indeed, we observe that the slopes are the same for different curves within a panel, which have the same el-ph coupling, whereas the slopes are steeper for full compared to reduced el-ph coupling. In fact, the ratio of extracted slopes is 0.8 and matches approximately the ratio of the bare coupling values $g^2$, despite the fact that the self-energies are computed self-consistently and thus contain also higher orders in $g^2$. The results clearly demonstrate the importance of el-ph coupling for the effective thermalization of the superconducting state at long times.

In order to gain additional insight into the interplay of collective order parameter dynamics and single-particle scattering during and after a 100 fs ramp, we show in Fig.~\ref{fig3} snapshots of the momentum-resolved dynamics of normal and anomalous densities, as well as the quasiparticle dispersions at selected times. We plot differences from the final thermal state, taken as the state with the final hopping value at the equilibrium temperature, to demonstrate the relaxation towards this state. 

Initially the distributions show strong deviations from the final thermal distribution in a broad momentum range. Right after the ramp, the normal distribution far from the Fermi surface quickly approaches the thermal distribution in both cases. Large deviations remain in a narrow region close to the Fermi surface. In contrast, the relaxation dynamics of the anomalous densities is much slower. In particular, for the case of smaller $\Delta_0$ almost no change can be detected, whereas for the intermediate $\Delta_0$ the momentum region of large deviations shrinks faster. Thus, the time scales for the relaxation of the normal densities are much faster than for the anomalous ones.

This is further supported by snapshots of effective Bogoliubov quasiparticle dispersions $E_\kk(t) = \sqrt{\tilde\epsilon(\kk,t)^2+\Delta(t)^2}$, compared with the thermal dispersions $E_{\kk,f} = \sqrt{\tilde\epsilon_f(\kk)^2+\Delta_f^2}$ (Fig.~\ref{fig3}(d-f) and (j-l)), where $\tilde\epsilon(\kk,t) \equiv Z \epsilon(\kk,t)$. Whereas for small $\Delta_0$ (upper panels), the deviation from $E_{\kk,f}$ is pronounced close to the Fermi surface, the dispersion for larger $\Delta_0$ (lower panels) is almost thermalized at 392 fs. 

\begin{figure}[ht!pb]
\includegraphics[clip=true, trim=0 0 0 0,width=\columnwidth]{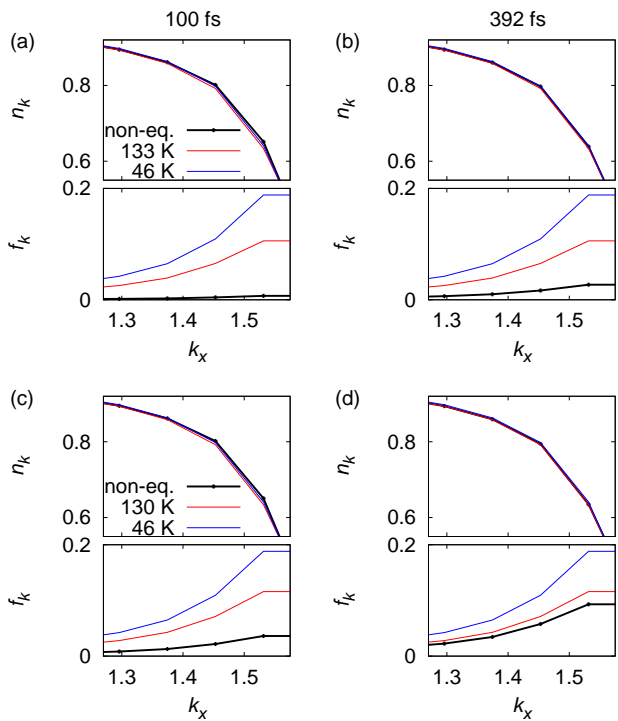}
\caption{
{\bf Cuts of momentum distributions.} Momentum distributions $n_k$ and $f_k$ along a diagonal cut $k_x = k_y$ for the same data as in Fig.~\ref{fig3} for $T=133$ K (panels (a) and (b)) and $T=130$ K ((c) and (d)), at times as indicated above. The black curves are instantaneous, non-equilibrium data, the red lines are final thermal distributions. Blue lines show thermal reference data for a lower temperature of 46 K.
}
\label{fig_wigner}
\end{figure}

In order to highlight the nonthermal character of the instantaneous distributions, we plot in Fig.~\ref{fig_wigner} the actual distributions compared with final thermal distributions as well as a reference set at much lower temperature of 46 K. Importantly, the out-of-equilibrium normal distribution $n_k$ correspond to a ``colder'' fictitious instantaneous temperature than both the final thermal and 46 K reference distributions. The anomalous distribution $f_k$, on the other hand, is too ``warm'' in all cases. This discrepancy demonstrates that an effective quasi-thermal discription of the non-equilibrium data as often used, e.g. in the two temperature model, is entirely inadequate here. This finding stresses the importance of a proper non-equilibrium modeling of the light-enhanced superconductivity.

\begin{figure}[ht!pb]
\includegraphics[clip=true, trim=0 0 0 0,width=\columnwidth]{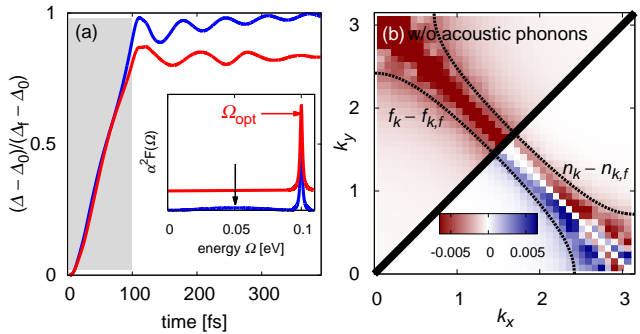}
\caption{
{\bf Importance of acoustic phonons for thermalization.} (a) Time-evolution of the order parameter in the presence of a narrow optic phonon mode (red curve) and additionally a broad acoustic phonon branch (blue curve). Inset: Eliashberg functions for the respective cases shifted for clarity. (b) Momentum-resolved deviations from final thermal values of normal and anomalous occupations at 392 fs in the presence of the narrow optic phonon mode (without acoustic phonons), shown in 1/8 of the Brillouin zone each separated by the black line. The relaxation processes are suppressed in a momentum window around the Fermi surface, which is determined by an energy window $E_{\kk}(\text{392 fs})$ $<$ $\Omega_{\text{opt}}/2$ (dashed curves).  
 }
\label{fig4}
\end{figure}

To emphasize the importance of the phonon spectrum, we compare in Fig.~\ref{fig4}(a) the order parameter dynamics with and without the low-energy acoustic phonon branch. The corresponding Eliashberg functions are shown in the inset to Fig.~\ref{fig4}(a). Clearly, the system effectively reaches the equilibration stage in the presence of the acoustic branch, whereas it is stuck at a nonthermal stationary value of the order parameter in the absence of the acoustic branch. As discussed before at long time scales in the presence of the acoustic phonons shows an exponential relaxation towards the thermal state with a time scale approximately proportional to $g^2$. 

To reveal the underlying reason for this nonthermal behavior in the abscence of acoustic phonons, we show in Fig.~\ref{fig4}(b) the momentum-resolved normal and anomalous density deviations from the final thermal values for the optical phonon. Clearly, there is a narrow window around the Fermi surface without allowed scattering phase space. This window is set by the optical phonon frequency and the Bogoliubov quasiparticle dispersion $E_{\kk}(\text{392 fs})$. A particle-hole pair with energy in the range $[-\Omega_{\text{opt}}/2, \Omega_{\text{opt}}/2]$ cannot relax because the required energy transfer is smaller than $\Omega_{\text{opt}}$. This provides a very intuitive explanation for the importance of acoustic phonons. Note, that the neglect of electron-electron scattering in our model is not the reason why the system shows this nonthermal behavior. In principle, it is correct that the thermalization of electrons amongst each other after excitation is facilitated by electron-electron interactions. However, this possible thermalization, in the absence of phonons, would be bound to occur at a higher effective temperature than the initial equilibrium temperature, simply by energy conservation in the closed electronic system. Thus, light-enhanced superconductivity would not profit from thermalization via electron-electron scattering.

\section{Conclusions and Outlook} 
\label{sec:conclusions}

In conclusion, we have demonstrated that nonequilibrium superconductivity can be enhanced on time scales reachable in pump-probe experiments with THz pump pulses. An enhanced electronic density of states around the Fermi level leads to strengthening of the effective pairing interaction, which dynamically enhances the superconducting order parameter during and after a ramp of the electronic hopping amplitude. The main features of the short-time dynamics are well described by a BCS model which means that the presence of the phonons mainly enters via the effective attractive interaction. In contrast, the presence of a phononic bath with a broad spectrum to which electrons can release energy is crucial to ensure that the thermal state is reached in which the superconducting order is fully enhanced to its expected equilibrium value after the ramp. Intriguingly, the phononic bath also enables the stabilization of the enhanced order parameter already for very fast ramps and thus opens an interesting route towards light-enhanced superconductivity on very short time scales.

The strong dependence of the dynamical enhancement of superconductivity on the initial order parameter raises the question of how to {\it induce} superconductivity when starting above $T_c$. A proper description of order parameter fluctuations, which trigger the symmetry breaking when starting in the normal state, is crucial in order to address this question (see Ref.~\onlinecite{warner_quench_2005} and references therein). Similarly, the nonequilibrium self-consistent update \cite{murakami_interaction_2015, schuler_time-dependent_2015, murakami_multiple_2015} of the pairing phonons is an interesting topic for future research.

\section{Acknowledgment}
We thank A.~Cavalleri, M.~Kollar, P.~van Loosdrecht, and A.~Subedi for discussions, and H.~van Pee for administrating the computer cluster on which the computations were performed. We acknowledge financial support of the DFG, and by the European Research Council (ERC-319286 QMAC, ERC-648166 Phon(t)on). 

\bibliography{Ramp}


\setcounter{figure}{0}
\setcounter{section}{0}
\setcounter{equation}{0}
\makeatletter 
\renewcommand{\theequation}{A\@arabic\c@equation}
\makeatother
\makeatletter 
\renewcommand{\thefigure}{A\@arabic\c@figure}
\makeatother
\makeatletter 
\renewcommand{\thesection}{A\@arabic\c@section}
\makeatother

\clearpage

\section{Appendix}
\label{sec:appendix}

\subsection{Time-dependent BCS equations}
\label{app:bcs}
As a simplified alternative to the full Migdal-Eliashberg theory, we also use time-dependent BCS-theory in order to describe the initial evolution of the system. The BCS Hamiltonian is given by
\begin{align}
\mathcal H &= \sum_{\bm{k}\sigma} \epsilon(\kk,t)  c^\dagger_{\bm{k}\sigma} c^{}_{\bm{k}\sigma}   \nonumber\\ &- |U| \sum_{\kk,\kk'} c^{\dagger}_{\kk' \uparrow} c^{\dagger}_{-\kk' \downarrow} c^{}_{-\kk \downarrow} c^{}_{\kk \uparrow},
\end{align}
where $U$ is the effective attractive interaction between the electrons, which is mediated by the electron-phonon interaction in the el-ph model. 

A mean-field decoupling in the Cooper channel leads to 
\begin{align}
\mathcal H &= \sum_{\bm{k}\sigma} \epsilon(\kk,t)  c^\dagger_{\bm{k}\sigma} c^{}_{\bm{k}\sigma}   \nonumber\\ &+ \sum_{\kk} \Delta c^{\dagger}_{\kk \uparrow} c^{\dagger}_{-\kk \downarrow} + h.c. + constants.
\end{align}

We define the normal and anomalous densities 
\begin{align}
n_\kk &= \langle c^{\dagger}_{\kk \sigma} c^{}_{\kk \sigma} \rangle,\\
f_\kk &= \langle c^{}_{-\kk \downarrow} c^{}_{\kk \uparrow} \rangle =f^{'}_\kk+if^{''}_\kk,   
\end{align}
with real ($^{'}$) and imaginary part ($^{''}$). The time-evolution equations for these densities are given by \cite{barankov_collective_2004, warner_quench_2005, yuzbashyan_nonequilibrium_2005, tsuji_theory_2015}
\begin{align}
\partial_t &f^{'}_\kk(t) = 2 \epsilon(\kk,t) f^{''}_\kk(t),\\
\partial_t &f^{''}_\kk(t) = -2 \epsilon(\kk,t) f^{'}_\kk(t) - \Delta(t) (1-n_\kk(t)-n_{-\kk}(t)),\\
\partial_t &\frac12(1-n_\kk(t)-n_{-\kk}(t)) = 2 \Delta(t) f^{''}_\kk(t).
\end{align}
The self-consistency condition is
\begin{align}
\Delta(t) &= -|U| \sum_\kk f^{'}_\kk(t),
\end{align}
where we have used that we chose the initial equilibrium solution to be real and given by
\begin{align}
\frac12(1-n_\kk(0)-n_{-\kk}(0)) &= \frac{\epsilon(\kk,0) \tanh(\frac{\beta E_\kk(0)}{2})}{2 E_\kk(0)},\\
f^{'}_\kk(0) &= \frac{-\Delta_0 \tanh(\frac{\beta E_\kk(0)}{2})}{2 E_\kk(0)},\\
f^{''}_\kk(0) &= 0,\\
1 &= |U| \sum_\kk \frac{\tanh(\frac{\beta E_\kk(0)}{2})}{2 E_\kk(0)},\\
E_\kk(0) &= \sqrt{\Delta_0^2 + \epsilon(\kk,0)^2}.
\end{align}
Here $E_\kk(0)$ is the well-known Bogoliubov quasiparticle dispersion that is obtained from the diagonalization of the mean-field BCS Hamiltonian, where $\Delta_0$ is the self-consistently determined initial equilibrium order parameter. 

The full solutions to the BCS equations are used in Fig.~2(a) and (c) of the main text. An analytical short-time solution for these equations can be obtained in the limit where we keep the normal density constant, $n_\kk(t) \approx n_\kk(0)$, and ignore self-consistent feedback by keeping $\Delta(t) \approx \Delta_0$ fixed. The short-time limit of the obtained solution is 
\begin{align}
f^{'}_\kk(t) - f^{'}_\kk(0) &= \frac{-\Delta_0 \tanh(\frac{\beta E_\kk(0)}{2})}{3 E_\kk(0)} \frac{\epsilon(\kk,0)^2 (J_0-J_f)}{J_0 \tau} t^3 \nonumber \\&+ \mathcal{O}((\epsilon(\kk,0) t)^4). 
\end{align}
This shows that for very short times $t < t_W$, smaller than the inverse of the initial electronic half bandwidth ($t_W =$ 0.658 fs for $W/2 = 4 J_0 =$ 1 eV), the change in the momentum-resolved order parameter within BCS theory scales cubically in time for all momenta. For times $t > t_W$ the short-time approximation breaks down and the self-consistent feedback from other momenta becomes crucial for the dynamics. Nevertheless, the extracted scaling with $\Delta_0$ shown in Fig.~2(b) of the main text shows that the initial order parameter sets the important dynamical time scale that governs the early-time enhancement of superconductivity during the ramp. 

\subsection{Intermediate time behavior}

\begin{figure}[ht!]
\includegraphics[clip=true, trim=0 0 0 0,width=\columnwidth]{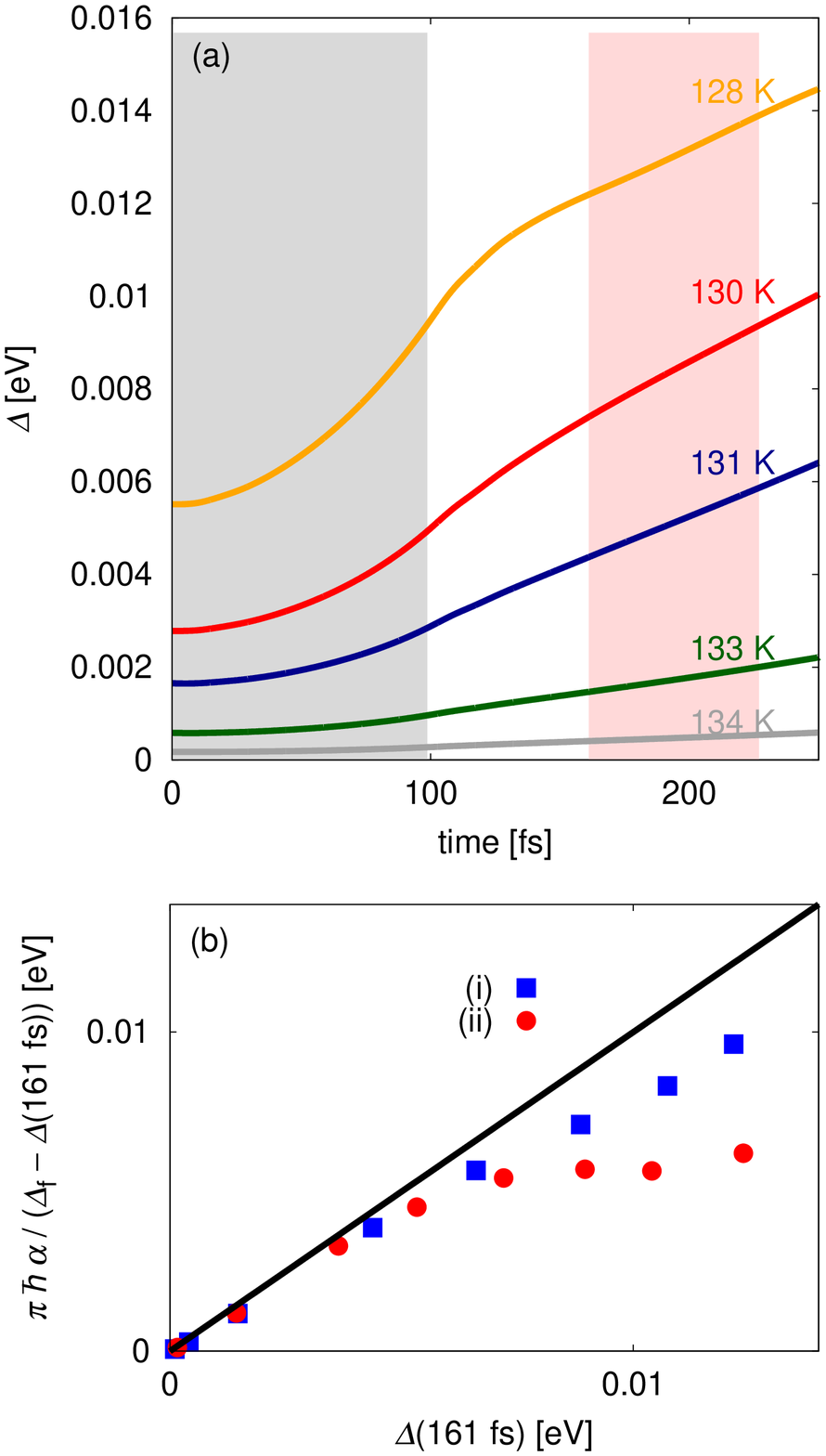}
\caption{
{\bf Intermediate time behavior.} (a) Time evolution of the order parameter for ramp duration $\tau = 100$ fs (grey shaded area). An approximately linear behavior is found for intermediate times (red shaded area). Here the temperature is varied as indicated, data are for the ``1.0 $g^2$''  el-ph coupling parameters. (b) Rates of change $\alpha$, obtained from fits to $\Delta(t)$ at intermediate times, scaled by remaining deviation $\Delta_f - \Delta(\text{161 fs})$, versus instantaneous order parameter. Colored arrows point to the parameter sets at $T =$ 133 K (magenta) and 128 K (orange). The black line indicates the equality between the scaled rate and instantaneous order parameter. 
}
\label{fig_SM}
\end{figure}

At intermediate times we observe a quasi-linear behavior of the time-evolution of the order parameter, if its value is still far from final thermal value. This regime is reached, if the ramp duration is short compared to $\Delta_0$, i.e.~if $\tau \Delta_0 < \hbar$.

Using linear fits to the intermediate time $\Delta(t)$ (see Fig.~\ref{fig_SM}(a))
\begin{align}
\Delta_{\text{fit}}(t) &= \Delta(\text{161 fs}) + \alpha(t-\text{161 fs}),
\end{align}
we obtain effective rates of change of the order parameter. In Fig.~\ref{fig_SM}(b) we show the rate normalized by the remaining deviations from the final thermal value, i.e.~the effective rate of change $\pi \hbar \alpha/(\Delta_f - \Delta(\text{161 fs}))$.  It is an interesting observation that the rates scale almost linearly with the instantaneous order parameter for small $\Delta(\text{161 fs})$. By contrast, the rates deviate from this linear behavior at larger $\Delta(\text{161 fs})$. In the regime of deviation, the rates also depend on the el-ph coupling strength. This behavior is in agreement with the observed momentum-dependent relaxation discussed in the context of Fig.~3 in the main text, and again illustrates the interplay of slow order parameter evolution far from thermalization and fast single-particle scattering becoming relevant for the order parameter dynamics close to thermalization. 

Note that the linear change of $\Delta(t)$ in the intermediate regime is an empirical observation in our full simulations for the electron-phonon system, and is for instance not matched by BCS results. It only holds for a certain temporal regime, and only for ``fast ramps'' with small initial $\Delta_0$, where the system is relatively far from its thermal value after the ramp. Thus, we identify heuristically a time scale for the increase of the order parameter at intermediate times, which will be important for experimental realizations. 

\end{document}